# Evidence of two different types of short term solar modulation of regional surface temperature and cloud.

I. R. Edmonds
12 Lentara St, Kenmore, Australia Q4069.
ian@solartran.com.au

**Abstract**
Recent work indicates that ~27 day variations in cosmic ray flux during 2007-9 are phase-locked to ~27 day variations in cloud and surface temperature at Shetland. Here we extend the study to other regions including Central England, US and Australia and to several other annual intervals that exhibit strong ~27 day variation in cosmic ray flux and sunspot area. Band pass filtering was used to obtain ~27 day components of daily maximum temperature in each region and ~27 components of cloud variation were determined, in Australia only, from solar exposure records. When cosmic ray flux is the dominant influence phase-locked variations in surface temperature occur in each of the regions with, however, in-phase or anti-phase variation in different regions. Similar phase–locking of ~27 day variation in surface temperature to sunspot area variation occurs when sunspot activity is the dominant influence with indications that changes from in-phase to anti-phase variation are linked to flipping of sunspot activity from one active longitude to another. The ~27 day component of cloud in Australia was phase-locked to the ~27 day component of temperature in Central England in two of the intervals studied, 1997 and 2005, indicating the global nature of the connection between solar disturbance and the lower atmosphere. It was observed that very large swings in ~27 day temperature components are often correlated with ~27 day variation in solar activity.

**1. Introduction**.
Studies of the connection between the variation of solar activity and the variation of surface temperature have been controversial, Gray et al (2010). It is generally accepted that there is a small but significant correlation between the strong ~11 year variation of solar activity and global surface temperature, Camp and Tung (2007), and that this correlation has a strong regional dependence, Lean and Rind (2008), Lockwood (2012). Solar activity also has a strong short term variation associated with the ~27 day rotation period of the Sun and it is well established that there are corresponding strong short term variations in cosmic ray flux, Alania et al (2008), and in the magnetosphere, thermosphere and upper atmosphere (Ruzmaikin et al 2007, Lei et al 2008, Liang et al 2008, Ma et al, 2012)  However, attempts to identify a corresponding short term variation of variables in the lower troposphere have been controversial. Perhaps the most controversial studies have been of the connection of short term variation in cloud (Svensmark et al 2009, Laken et al 2011, Calogovic et al 2010) and surface temperature, Dragic et al (2011), to the short term variations in cosmic ray flux known as Forbush decreases. The essential problem with solar activity influencing the variability of tropospheric temperature is that the energy change available directly from variation in solar activity is about 1 W/m$^2$ in 1380 W/m$^2$ or about 0.1% and is so small that any direct effect of short term solar activity would be orders of magnitude below the noise level of temperature variations due to synoptic weather change. Relatively small changes in cloud cover have a strong effect on surface temperature hence the interest in the possibility of solar activity influencing cloud formation and, in turn, temperature. Sophisticated analysis of data records have been able to detect small ~27 day components in some tropospheric variables, e.g. vorticity, Wilcox et al (1974), potential height, Burns et al (2008), cloud, Takahashi et al (2010) and potential gradient, Harrison et al (2011). A further problem is that the forms of solar activity relevant to short term variability occur irregularly, e.g. sunspots, coronal holes, Forbush decreases, so that conventional methods of detecting weak periodic variations in long records may not be efficient in this case e.g. Brown (2008). However, recently Harrison et al (2013) have used a band pass filter method to establish a connection between ~27 day cosmic ray flux variation, cloud base height and surface temperature during an interval around January 2008.

Establishing a deterministic relationship between two variables is difficult when the variables are complex and chaotic but ultimately related. It is necessary to eliminate the possibility that the two variables may be unrelated, random processes, Lockwood (2012). Distinguishing between the two possibilities is especially difficult when the proposed deterministic process, for example the dependence of surface temperature on solar activity, may be due to several different deterministic processes and occurs, intermittently, in combination with random processes such as synoptic variation in local weather. The usual approach is to identify some characteristic of the deterministic process e.g. a characteristic frequency by means such as frequency analysis or a characteristic time variation by epoch superposition. This process is simplified if a linking mechanism which predicts the characteristic frequency or the characteristic time variation is known. Conventionally, the significance of any relationship uncovered is estimated by comparing the identified frequency variation or time variation in the variable of interest with the frequency variation or time variations in generated red or white noise of similar variance. Torrence et al (1998). However, the estimation of significance can often be controversial e.g. Le Mouel et al (2010) and Legras et al (2010). Another common, but less formal, approach is to select a short time interval from a long record when one variable exhibits a well defined time variation or frequency component and examine if the dependent variable exhibits a similar time or frequency dependence. For example, if one identifies a strong 27 day component in cosmic ray flux during some interval one might examine if the Sun exhibits a single, large coronal hole during the same interval, Rouillard and Lockwood (2007), or if cloud exhibits a similar periodicity, Harrison et al (2011). However, this type of approach is open to criticism on the basis of selection bias in the sense that such correlation could have arisen by chance in long noise records, Lockwood (2012). The significance of this approach can, therefore, be increased if the occurrence of one of the variables has been examined in independent studies thus avoiding any suggestion of selection bias with respect to time interval. This is the approach adopted here. We examine time intervals when independent observers have reported on characteristics of our independent variable, in the present case, solar activity. We then examine whether our dependent variables, surface temperature or cloud, exhibit similar time and frequency characteristics to the independent variable. The publication used here is "Prediction of solar active longitudes" Zhang et al (2008). This reports well defined times of emergence of active regions on the Sun during two years, 1997 and 2005, and provides a clear time signature of solar activity in each of the 1997 and 2005 years. During 1997 numerous X-ray flares occurred on two active longitudes $180^o$ apart; the first group occurred between day 189 and day 230 and the second group occurred between day 270 and day 300. During 2005 a group of X-ray flares occurred on one active longitude between day 1 and day 95 and on the other active longitude between day 135 and day 257. It turns out that there are other studies of the Sun-Earth connection during the years 1997 and 2005, (Rouillard and Lockwood 2007, Harrison et al 2011, Lei et al 2008, Ulrich et al, 2002), that add further data associated with the two intervals. Harrison et al (2011), using a band pass filter method similar to the method used in this paper, were able to demonstrate an extended interval of correlated ~27 day variation of cosmic ray flux and atmospheric potential gradient in 1996 and 1997 and related these variations to short term variation in cloud base height. Lei et al (2008) reported strong ~27 day periodicities in the ionosphere electron content in 2005 and Ulrich et al (2002) provided a detailed record of short term variation in solar magnetic field (super-synoptic charts) covering the 1996 – 1997 interval. So in 1997 and 2005 there are well defined time signatures of sunspot and flare activity as identified by Zhang et al (2008) and other studies that indicate the presence of persistent coronal holes.

## 2. Method and data sources

We use a method of band pass filtering similar to that used by Harrison et al (2011) and Harrison et al (2013) to identify ~27 day components in cosmic ray flux and potential gradient, by Lean and Rind (2008) to compare the ~11 year components in surface temperature with solar activity and by Ma et al (2012) to compare the ~27 day components in $F_2$ region peak electron densities with solar activity. In the present study the dependent variables are daily maximum surface temperature, T, and daily average cloud fraction. Isolation of the ~ 27 day components of the variables was obtained by making a Fast Fourier Transform of each data series. The resulting n Fourier amplitude and phase pairs, $A_n(f_n)$, $\phi_n(f_n)$, in the frequency range 0.031 days$^{-1}$ to 0.045 days$^{-1}$ (period range 32 to 22 days) were then used to synthesize a band pass filtered version of each variable, denoted for example 27T, by summing the n terms, $T_n = A_n Cos(2\pi f_n t - \phi_n)$ for each day in the series. Where a data series has been smoothed by, for example, a 30 day running average, the resulting smoothed series is denoted e.g. T S30.
Central England temperature data, TCET, was obtained from
http://www.metoffice.gov.uk/hadobs/hadcet/cetmaxdly1878on_urbadj4.dat. Melbourne temperature data, TMEL, was obtained from the Australian High-Quality Climate Site Network, http://www.bom.gov.au/climate/change/hqsites/. Boulder temperature data, TBOU, was obtained from
http://www.esrl.noaa.gov/psd/boulder/data/boulderdaily.complete .
Sunspot area data, SSA, was obtained from
http://solarscience.msfc.nasa.gov/greenwch/daily_area.txt. Cosmic ray flux data, CRF, was obtained from ftp://ulysses.sr.unh.edu/NeutronMonitor/DailyAverages.1951-.txt .
Australian solar exposure data, E, was obtained from
http://www.bom.gov.au/climate/data/ . Cloud cover data for Hobart was purchased from the Australian Bureau of Meteorology.

## 3. Comparison of ~27 day sunspot area and surface temperature variations.

Daily sunspot area has been recorded by Greenwich Observatory since 1875 and is ideal for examining short term variation in solar activity associated with sunspots and flares. In Figure 1A we show that the observed variation of sunspot area during 1997 and 2005 is consistent with the occurrence reported by Zhang et al (2008) of X-ray flares in 1997 and 2005. That is, there are two strong bursts of sunspot area in the latter half of 1997, Figure 1A, and the occurrence of strong bursts of sunspot area extending over the entire interval of 2005, Figure 1B. As, in this paper, we will be comparing ~27 day variations in solar activity to ~27 day variations in surface temperature and cloud, the time axes in Figure 1A and 1B, and in most subsequent Figures, are divided into intervals of 27 days to facilitate following phase change of a variable relative to solar rotation and phase changes between variables.

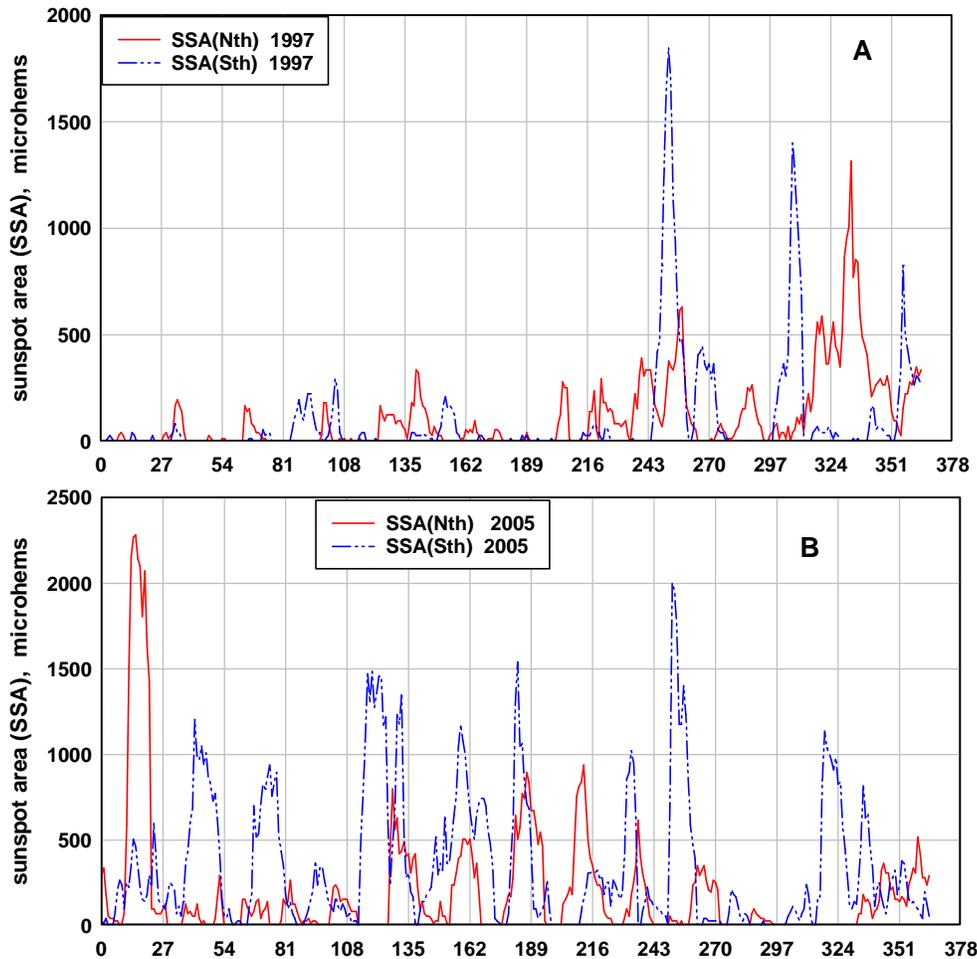

**Figure 1.** Variation of sunspot area North and sunspot area South during (A) 1997 and (B) 2005.

The raw data of sunspot area and Central England temperature are compared in Figure 2A. The seasonal variation of TCET and variations of period longer than 30 days have been removed from the raw data by subtracting the 30 day moving average. As shown in Figure 2A, the ~27 day periodicity in solar activity, in this case SSA(Nth), is usually evident in the raw data of quantities like sunspot area, sunspot number or cosmic ray flux. However, the ~27 day component in surface temperature is usually much less evident and needs to be separated from synoptic weather noise by band pass filtering. Figure 2B compares the ~27 day components, obtained by band pass filtering, of sunspot area North, 27SSA(Nth), and daily maximum temperature in Central England, 27TCET, in 2005. For year 2005 there is an excellent correspondence between the amplitude of the 27SSA(Nth) and 27TCET variation during the middle of the year. The phase difference between 27SSA(Nth) and 27TCET changes slowly from $180^o$ to $90^o$ during this interval.

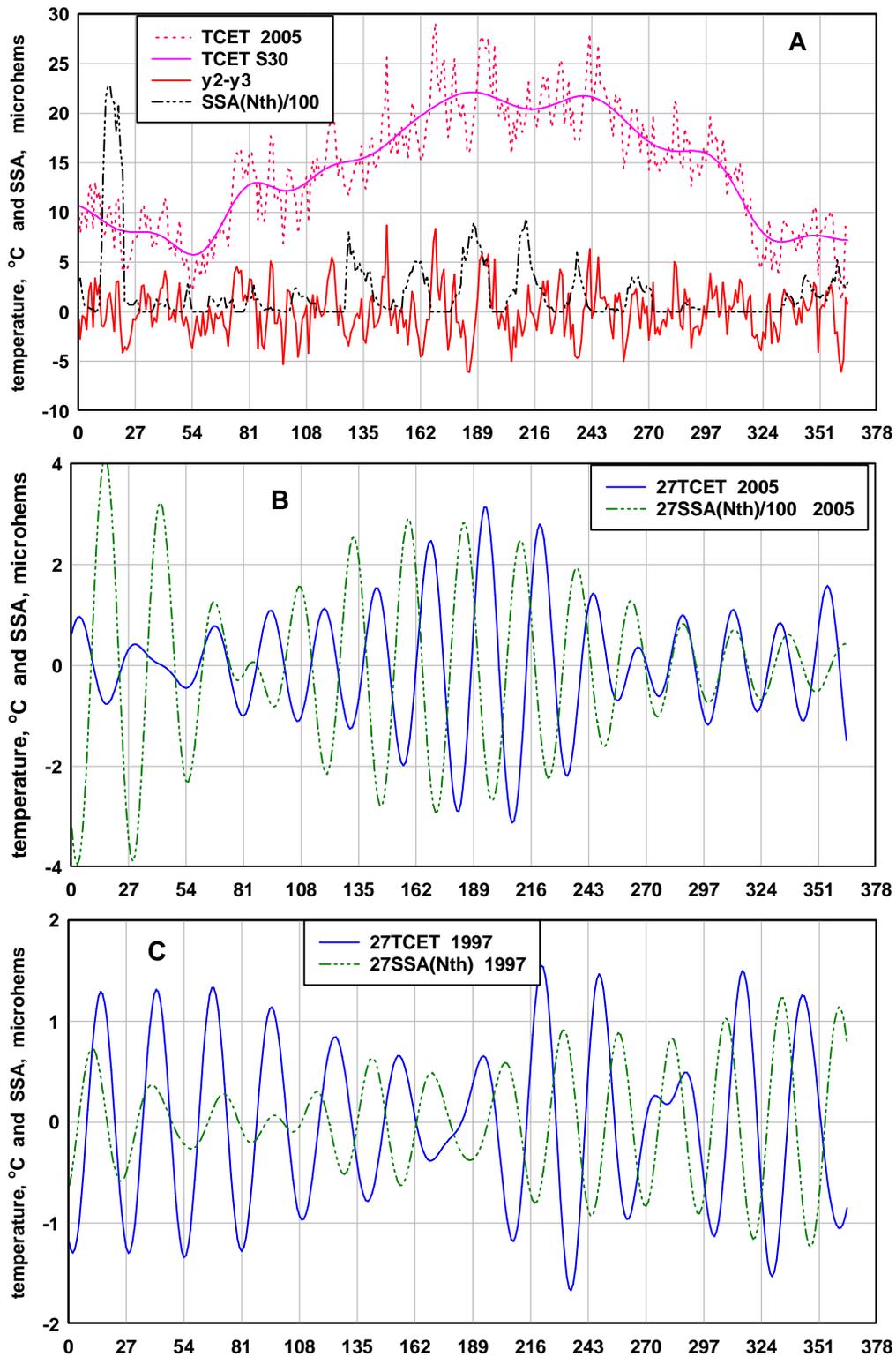

**Figure 2.** (A), compares the variation of sunspot area (North) with the short term variation of Central England temperature (TCET) in 2005. (B) Compares the 27 day components, 27SSA(Nth) and 27TCET, in 2005. (C) Compares 27SSA(Nth) and 27TCET in 1997.

There will be considerable focus in this paper on phase change between the various components examined and the phase change of individual components relative to solar rotation – as indicated by the 27 day spacing of the time axis. For example there is clearly a $180^o$ phase change, (a time shift of about 13.5 days), in the variation of 27SSA(Nth) between early 2005 and the middle of 2005 that is consistent with the shift of solar activity from one active solar longitude to the other active longitude on the opposite side of the Sun (Zhang et al, Fig 2, 2008). Figure 2C compares 27SSA(Nth) and 27TCET for 1997. The ~27 day component of temperature increases in amplitude during the time intervals corresponding to high sunspot and flare activity as indicated by the observations of Zhang et al (2008) for 1997 (day 189 to 230 and day 270 to 300). However, there are significant differences. First, the time intervals for bursts of X-ray flares as indicated by Zhang et al (2008) are shorter, by about 2/3, than the two time intervals of high 27TCET in the second half of 1997. However, this may be an artefact of the band pass filter method which delivers a wider variation, (a sinc function) to short term, (impulsive) time variations. During the first half of 1997 there is a high 27TCET component when there is low solar flare activity according to Zhang et al (2008) and low sunspot area, according to Figure 1A. This high 27TCET component at the beginning of 1997 is almost certainly associated with the presence of an isolated equatorial coronal hole and the corresponding ~27 day variation in cosmic ray flux observed by Harrison et al (2011). Also, whereas the phase of the 27SSA(Nth) variation in 1997 is relatively constant with respect to 27 day solar rotation the phase of the 27TCET variation suffers a $180^o$ phase change between day 162 and day 189 and again between day 270 and day 297. The phase of 27SSA(Nth) relative to solar rotation is stable throughout 1997 so the $180^o$ shift in phase of 27TCET between day 162 and day 189 and between day 270 and day 297 must be due to some other, possibly terrestrial effect or possibly a two mechanism effect, as discussed later. Thus, while the result for 2005 suggests a simple relation between sunspot area and surface temperature the result for 1997 suggests a more complex relation in which the influence of a coronal hole in the early part of 1997 overlaps the influence of sunspot and flare activity in the last half of 1997.

The observed ~27 day component of surface temperature is quite large: about $1.5^oC$ amplitude in 1997 and up to $3^oC$ amplitude in 2005. Such large variations in temperature cannot be due, directly, to changes in irradiance associated with solar activity. The changes in total solar irradiance due to variation in sunspots over an 11 year solar cycle amount to about 1 $W/m^2$ in 1370 $W/m^2$ or about 0.1% and do lead to a variations in global surface temperature of about $0.1^oC$, (Lean and Rind 2008, Camp and Tung 2007). The ~27 day variation in total solar irradiance due to long lived sunspots is also of the order 1 $W/m^2$, Svaalgard (2008), and should also directly lead to ~27 day variations of about $0.1^oC$ in daily maximum temperature, variations that would be completely hidden in synoptic noise. Therefore ~27 day variations of the order of $3^oC$ amplitude, Figure 2, must be due to some indirect process. The variation of cloud due to variation in cosmic ray flux is one candidate process and a process that has received much attention (Svensmark and Friis-Christensen 1997, Rycroft et al 2008, Tinsley and Yu 2004, Kirby 2007, Harrison et al 2011, Harrison et al 2013). However, there are other candidate solar influenced processes that could lead to large regional temperature variations e.g. solar induced perturbations of the jet streams (Huth et al 2008, Woolings et al 2010 and Lockwood 2012) and variations in the jet streams due to stratospheric warming, Haigh (2003). Here, due to the ready availability of daily cosmic ray flux data we focus on examining correlations of cosmic ray flux, cloud and temperature. The basis of the cosmic ray flux - cloud mechanism is that solar activity e.g. coronal holes or coronal mass ejections transport the Suns magnetic flux into the heliosphere and this magnetic flux shields the inner heliosphere and Earth from galactic cosmic rays. This shielding process is a complex, poorly understood, charged particle diffusion mechanism, Alania (2008). Cosmic rays may act in several ways to alter cloud formation in the atmosphere, (1), by seeding the aerosol to cloud growth mechanism or (2), by altering the global electric circuit and varying the formation of clouds via space charge effects at cloud boundaries, (Tinsley et al 2007, Rycroft et al 2008, Yu 2002, Tinsley and Yu 2004, Rycroft et al 2012, Svensmark et al 2013, Nicoll and Harrison 2013). Neither process is well understood and the studies of the effect of cosmic rays on the global electric circuit and clouds have been limited to short intervals (Svensmark et al 2009, Harrison et al 2011, Harrison et al 2013) and have been controversial, Calogovic et al (2010). However, a small change in cloud cover can, in principle, lead to relatively large changes in solar and long wave exposure of the surface and in surface air temperature.

**4. Comparison of ~27 day cosmic ray flux and surface temperature variation.**
Cloud influences surface temperature by reducing exposure of the surface to solar radiation and by trapping long wave radiation. The most basic expectation is: higher cosmic ray flux (CRF) leads to more cloud, less solar exposure and lower surface temperature. However, this basic idea is complicated as clouds also absorb long wave radiation from the surface and reradiate it downwards as well as upwards. Low cloud is warm and effectively emits long wave radiation to space. Therefore, the dominant effect of low cloud is cooling. High cloud is cold and tends to trap long wave radiation rather than emitting it to space. Therefore the net effect of high cloud is a warming effect. So the more detailed possibilities are: if higher CRF leads to more high cloud the surface warms whereas if higher CRF leads to more low cloud the surface cools. A further complication may arise from the different effect of CRF on high and low cloud. Yu (2002) and Tinsley and Yu (2004) have calculated that CRF tends to reduce high cloud and increase low cloud. So a further possibility is: higher CRF reduces high cloud and the surface cools and higher CRF increases low cloud and the surface cools. Thus the effect of the CRF – cloud - solar exposure - surface temperature mechanism is complicated with several competing effects. A further complication arises from the possibility of a significant time delay, several days, between a peak or minimum in CRF and the resulting effect on cloud, solar exposure and surface temperature (Svensmark et al 2009, Kirby 2007). However, recent studies by (Harrison et al, 2013, Nicoll and Harrison 2013) indicate that the response of cloud and temperature to cosmic ray change is almost instantaneous on the time scale, 27 days, used in this work. It can also be anticipated that the short term relationship between CRF, cloud and surface temperature will be dependent on the presence and altitude distribution of water vapour and cloud, both of which vary seasonally and daily with synoptic weather variation. Filtering to obtain ~27 day variations will average synoptic cloud and temperature variation to some extent but not seasonal cloud and temperature effects. Thus the phase relationship between CRF and temperature is expected to be complicated by seasonal effects.

Figure 3A shows the percentage variation of CRF and 27CRF during 1997 with the 27TCET variation superimposed. The phase and amplitude of 27CRF relative to solar rotation is remarkably stable right through 1997 and it is interesting that during the two intervals of high solar activity in the latter half of 1997 the 27CRF variation tends to be in phase with the several large Forbush decreases in CRF that occur at days 270, 297 and 324. The fact that the 27CRF phase remained stable during the intervals of intense sunspot and flare activity may be due to the fact that the coronal mass ejections leading to the in-phase Forbush decreases were effective because they were directed towards Earth and their effect therefore in phase with solar rotation. Over the first six solar rotations in first half of 1997 the ~27 day components of CRF and TCET are in-phase. About day 189, 27TCET undergoes a $180^{o}$ phase change then, at about day 270, reverses the phase change to be in phase with 27CRF again. While the phase changes occur at the start of each interval of high sunspot and flare activity in 1997 the fact that the phase of the 27CRF variation remains stable with respect solar rotation suggests that there may be two different mechanisms influencing surface temperature – one associated with coronal holes and cosmic ray flux and one associated with sunspot activity.

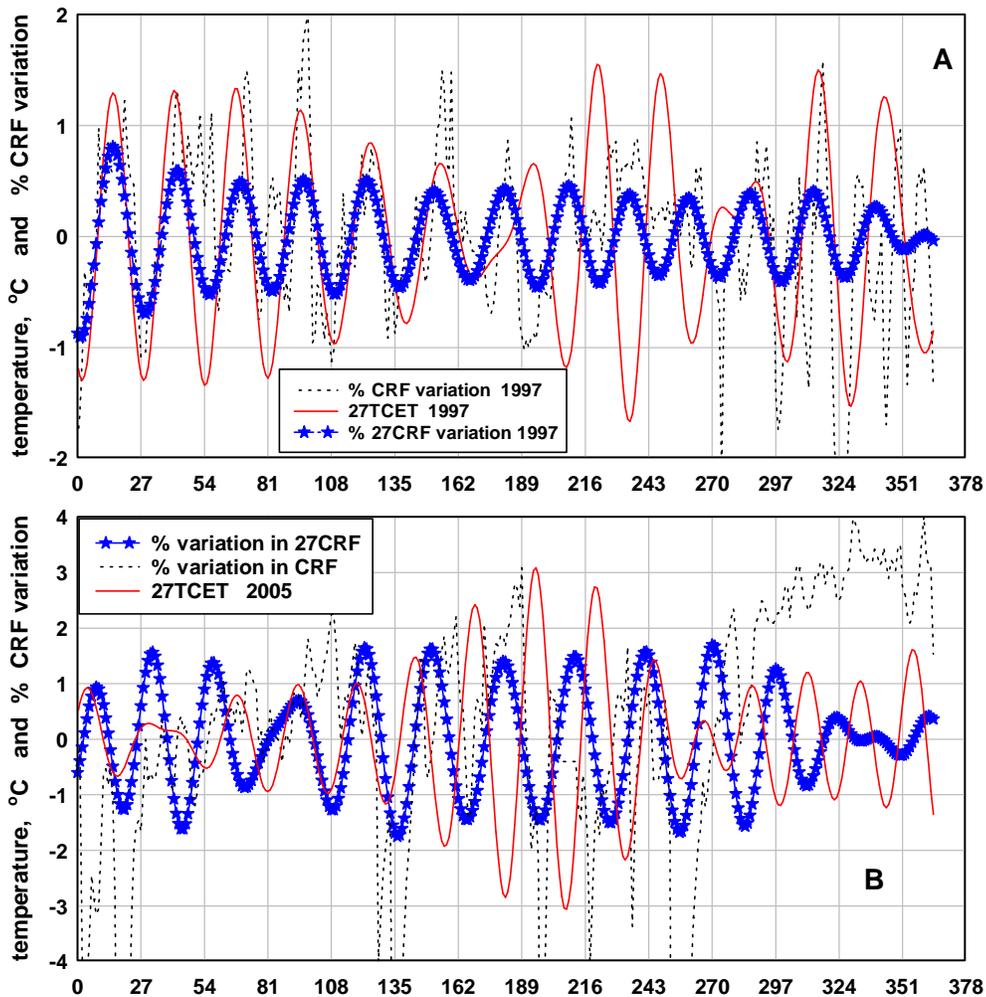

**Figure 3.** Compares the percentage variation of cosmic ray flux (%CRF), the ~27 day component of cosmic ray flux (%27CRF) and the ~27 day component of Central England temperature (27TCET) in (A), 1997 and (B) 2005.

In 2005 the variation in CRF is dominated by large Forbush decreases, at day 20, 135, 200 and 250, Figure 3B, and it is clear that during 2005 the CRF variation must have very significant components at periods shorter than 27 days. However, it is known that a coronal hole existed in 2005, Lei et al (2008), and may also have had some influence. The 27CRF variation that is observed exhibits the $180^o$ phase change at around day 81 that corresponds to the phase change in SSA(Nth) noted previously in Figure 2B that was associated with the change between active longitudes observed by Zhang et al (2008). The phase relationship between 27CRF and 27TCET is complicated but the components are mainly anti-phase in the middle of the year when the 27TCET response is maximised. This is consistent with the most basic form of CRF – cloud – temperature variation: increase in CRF leading to more cloud and lower surface temperature.

The interval in the first half of 1997 is interesting as it is known, Zhang et al (2008), that during the first half of 1997 there is little or no flare activity on the Sun and, from Figure 1A it is known that there is little sunspot activity on the Sun. Therefore during this time it is expected that solar wind velocity and therefore the variation in CRF will be dominated by the tilt of the heliospheric current sheet and the distribution of solar coronal holes. It is known that a long lived coronal hole near the Suns equator persisted through 1996 into 1997, Rouillard and Lockwood, (2007), and this coronal hole was effective in producing a long interval of ~27 day variation in CRF, Harrison et al (2011). The persistence of this coronal hole and the ~27 day variation of CRF into the first half of 1997 is clearly indicated in Figure 3A. Of particular interest for this paper is that, in the first half of 1997, 27TCET is in-phase with 27CRF.

Figure 4A shows that the in-phase correlation between 27CRF and 27TCET extends back into 1996 until about autumn equinox (day 287 reference line) when a phase change of $180^o$ in 27TCET occurs. Other $180^o$ phase shifts of temperature relative to cosmic ray flux occurred near day 544 and day 649, corresponding to the times of emergence of flare and sunspot activity previously noted. Using essentially the same band pass method as used here, Harrison et al (2011) demonstrated that for about a quarter of a year on either side of January 01 1997 the ~27 day components of CRF and atmospheric potential gradient were in-phase. It is interesting that over much of this same interval the ~27 day component of TCET is in-phase with the ~27 day components of CRF and atmospheric potential gradient. This in-phase variation of temperature with cosmic ray flux is the opposite of the simple expectation of more CRF, more cloud and lower temperature. A similar coronal hole and CRF variation occurred in the 2007 2008 interval, Harrison et al (2013). Figure 4B compares the ~27 day components of cosmic ray flux and Central England temperature during this period. It is evident from the CRF variation that the coronal hole emerged about day 254 and remained in stable phase relative to solar rotation through to day 702. However, the ~27 day temperature variation was in-phase with the CRF variation between day 254 and 386, out-of-phase between day 386 and 468 and in quadrature phase until near the end of 2008.

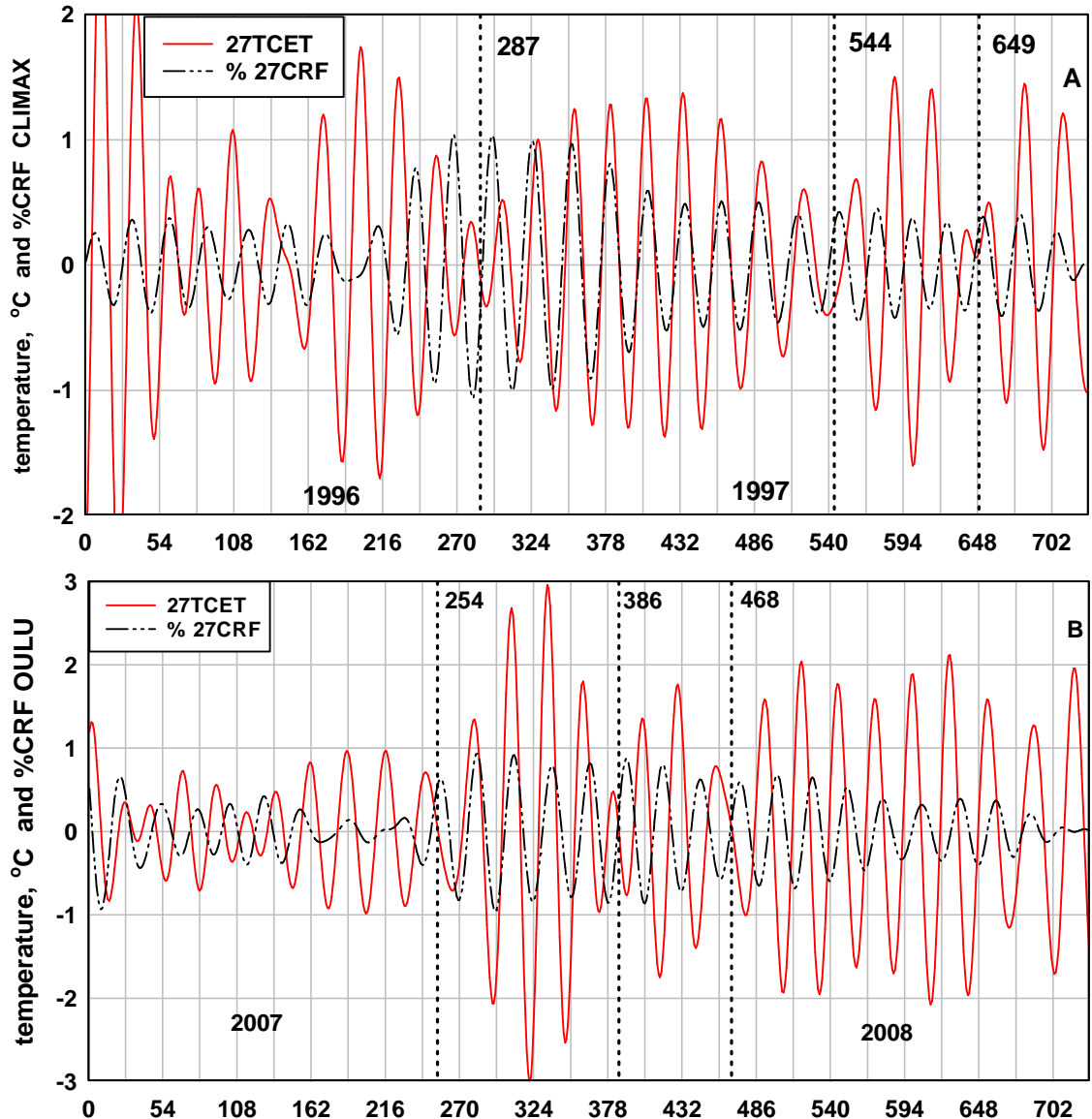

**Figure 4.** Comparison of the ~27 day components of cosmic ray flux and Central England temperature, 27TCET, during (A), 1996 and 1997 and (B), 2007 and 2008.

## 5. Comparison of ~27 day temperature variations in northern and southern hemisphere regions.

If a solar activity – CRF – cloud – surface temperature process occurs, the expectation is that it would be complicated by seasonal difference in the southern and northern hemispheres but should occur globally. That is, if there is a ~27 day variation in the northern hemisphere there should be a corresponding, but not necessarily in-phase ~27 day variation, in the southern hemisphere. Here we compare the surface temperature variation at Melbourne, Australia, with the temperature variation at Central England during 1997 and 2005. As Melbourne and Central England are nearly antipodal it is expected that the synoptic weather at Melbourne should be unrelated to the synoptic weather at Central England and that any common component must be due to an external, global, influence.

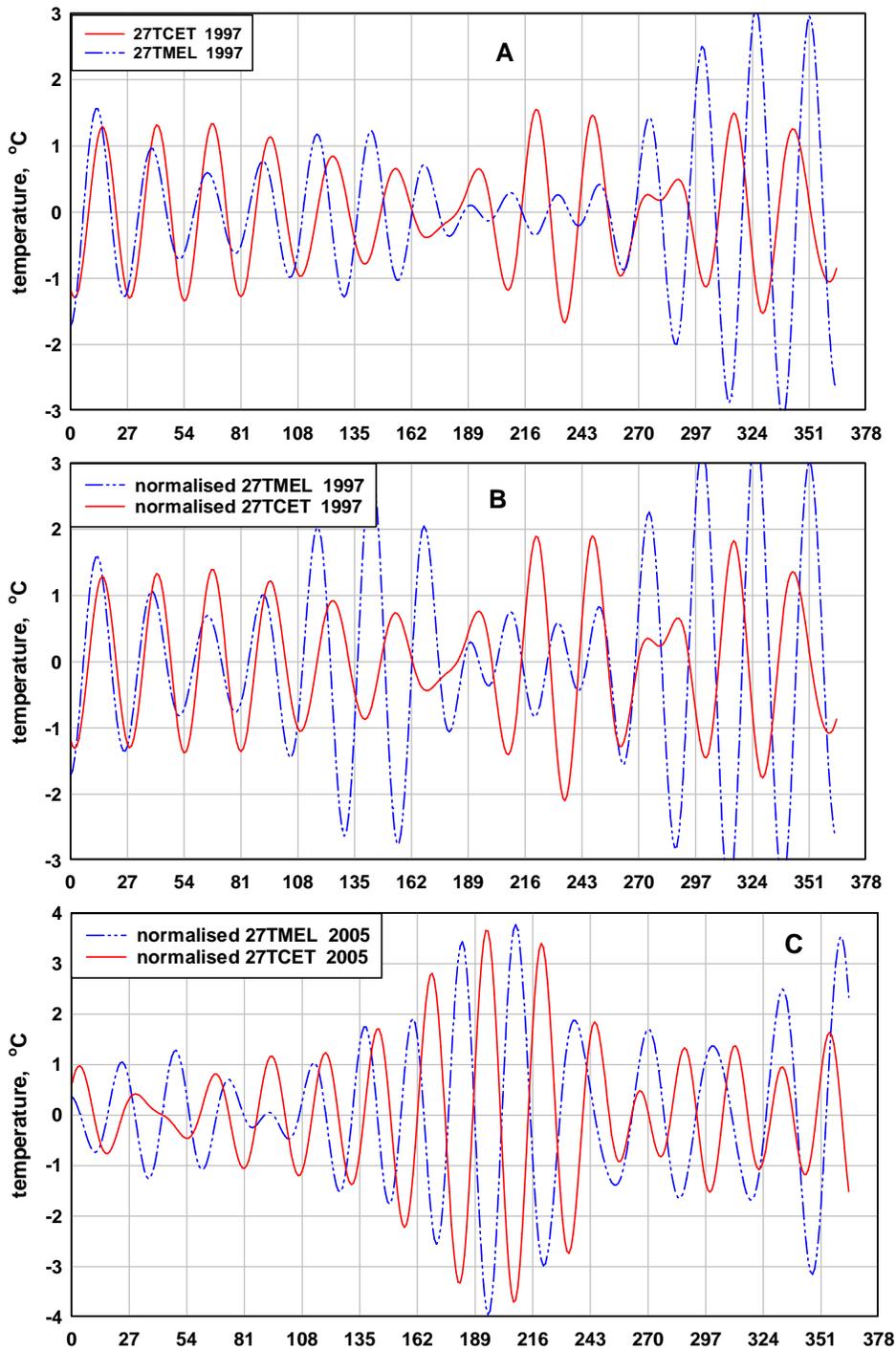

**Figure 5.** (A) The ~27 day components Central England and Melbourne temperature, 27TCET and 27TMEL, in 1997. (B) The ~27 day components, 27TCET and 27TMEL, in 1997 after normalisation against average seasonal variability. (C) The normalised variation of 27TCET and 27TMEL in 2005.

Figure 5A shows 27TCET and 27TMEL for 1997 and there are some common features. However, there is a considerable difference in the seasonal variability of surface temperature at the two locations that may be relevant in making comparisons. Melbourne is on the coast between the dry interior of Australia and the cold Southern ocean. As a consequence it suffers high seasonable variability in surface temperature due to synoptic weather patterns. The short term temperature variability in summer is about two times the temperature variability in winter. The surface temperature variability in Central England is relatively more constant. Adjustment for seasonal variability can be made by obtaining the average daily standard deviation of TCET and TMEL over a number of years and normalising the variation in a single year by dividing the daily variation in that year by the average daily variation to obtain a normalised variation. Figure 5B and 5C compare the normalised ~27 components of TCET and TMEL for 1997 and 2005.

In 1997 27TMEL and 27TCET are in-phase during the early part of 1997 but thereafter the relationship is primarily out-of-phase. In particular 27TMEL does not respond as significantly as 27TCET to the occurrence of flare activity and sunspots between day 189 to 230. Rather, by reference to Figure 2C, we see a strong, in-phase, response to the 27SSA(Nth) variation that occurred between day 108 and day 189 and between day 270 and day 378. For 2005, Figure 5C, there is a clear anti-phase relationship between 27TMEL and 27TCET during the middle of the year. With reference to Figure 3B it is clear that during the middle of 2005, i.e. during winter in Melbourne, the 27 day component of surface temperature is varying in-phase with the 27 day component of cosmic ray flux. These results provide tentative support for the concept of a seasonal effect resulting in anti-phase variations of temperature in opposite hemispheres when influenced by sunspot and flare activity but in-phase variations in temperature when influenced by cosmic ray flux variation associated with coronal holes.

Harrison et al (2011) observed an in-phase ~27 day variation in CRF and atmospheric potential gradient extending from the middle of 1996 and well into 1997. Roulliard and Lockwood (2007) showed that this variation was due to a single persistent and geo-effective coronal hole. Figure 4A showed that 27TCET was in-phase with 27CRF over much of this interval. Figure 6 compares 27TCET and 27TMEL during 1996 and 1997 and shows that the two temperature components are in-phase for 300 days or 12 solar rotations over the same interval of strong 27CRF flux variation (Figure 4A) and strong ~27 day potential gradient variation, Harrison et al (2011). This suggests there is a relatively simple connection between the existence of a single geo-effective equatorial coronal hole and cosmic ray flux, potential gradient and surface temperature at antipodal locations. If we focus on day 366 in Figure 6 the observed connection is a decrease in 27CRF, see Figure 4A, leading to a decrease in both 27TCET and 27TMEL. Later in this paper we show that, at this same time, the decrease in CRF leads to an increase in cloud at Melbourne – a somewhat counter intuitive result. It seems unlikely that, coincident with a long interval of 27CRF variation, a 300 day long interval of strong, in-phase, ~27 day variations in surface temperature would be observed at antipodal locations if the temperatures at the antipodal locations were unrelated random variations of synoptic weather.

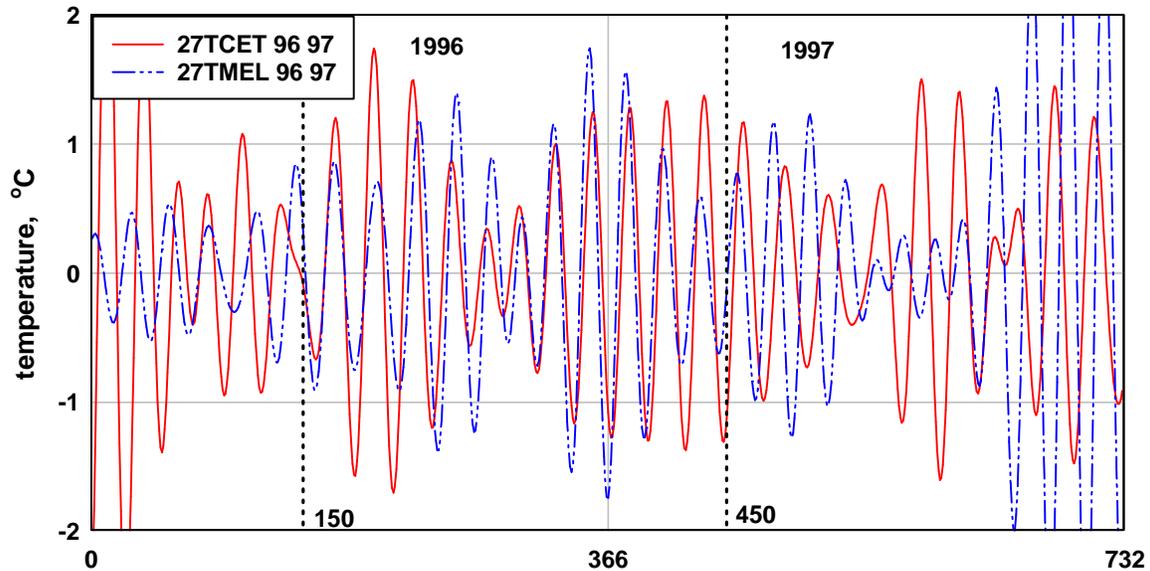

**Figure 6.** Compares the ~27 day components of Central England and Melbourne temperature during 1996 and 1997.

**6. Comparison of cosmic ray flux, cloud fraction and temperature variation.**

There are numerous studies on the connection between galactic cosmic ray and cloud variations. The best known and possibly most controversial are studies to observe the 11 year solar cycle variation of global cloud cover with cosmic rays, Svensmark and Friis-Christensen (1997) and a short term decreases in global cloud with Forbush decreases in cosmic rays, Svensmark et al (2009). Harrison (2008), Harrison and Ambaum (2010), Harrison et al (2011) and Harrison et al (2013) have studied the connection between short term CRF variation and cloud in England. However, these studies were hampered by limited cloud data and relied on either measures of diffuse daylight fraction or cloud base height. In Australia there is an extensive network of stations recording daily solar exposure, E. In the absence of cloud the daily solar exposure, $E_0$, varies in a manner that depends on transmission of the atmosphere and declination of the Sun and is, therefore, relatively simple to calculate using solar geometry, Duffie and Beckman (2006), and basic clear sky models of horizontal global irradiance such as the Haurwitz model, Haurwitz (1945), Reno et al (2012). The ratio, $E/E_0$, represents the daily average transmission through the cloud cover and $(1 – E/E_0)$ is the daily average cloud reflectance and can be taken as a measure of the daily average cloud fraction.

Figures 7A and 7B show the cloud fraction and ~27 day component of cloud fraction for 1997 and 2005 calculated from the solar exposure data from the station at Melbourne Airport South Anemometer, (note: the last two months of solar exposure data in 2005 are missing from all the recording stations in Australia). A noticeable feature is that the ~27 day variation of cloud fraction in Melbourne has an amplitude variation that follows the time signature of the solar activity events for 1997 and 2005 discussed earlier. In particular increased amplitude between days 216 and 270 and days 297 and 351 of 1997 that correspond to the intervals of sunspot and flare activity and, in the early part of 1997, the variation associated with a persistent coronal hole. By comparing 27TMEL in Figure 5A with $27(1 – E/E_0)$ in Figure 7A it is seen that surface temperature and cloud fraction vary in almost exact anti-phase through 1997 except for the interval between day 189 and day 270 when the variation is in-phase and when increasing cloud leads to increasing temperature.

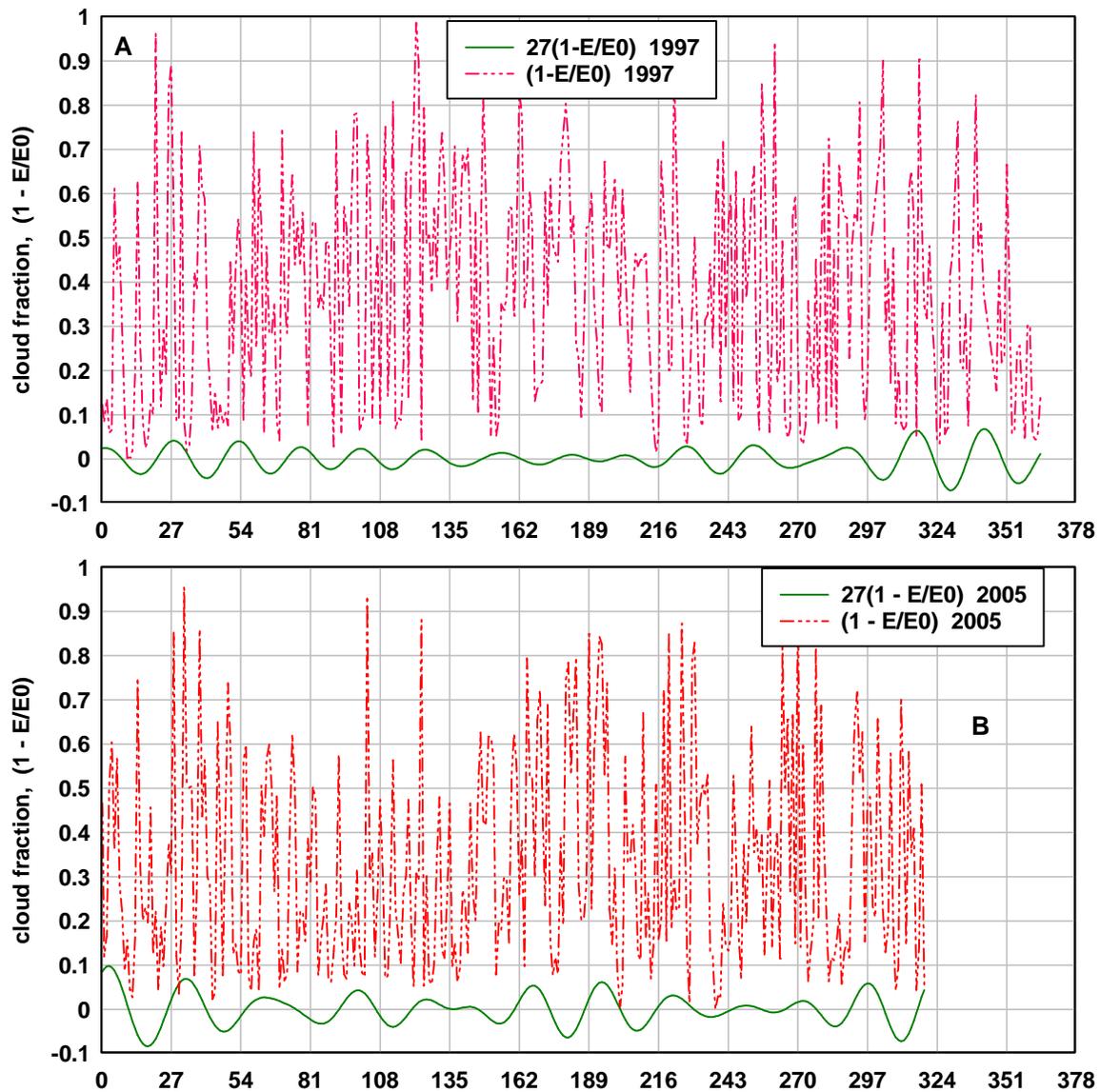

**Figure 7.** The cloud fraction at Melbourne, $(1 - E/E_0)$, as determined from global horizontal solar exposure with cloud, $E$, and when cloud free, $E_0$, along with the ~27 day component of cloud fraction, $27(1 - E/E_0)$ for (A) 1997 and (B) 2005. The phase and amplitude of the ~27 day component of cloud fraction at Melbourne can be compared with the phase of ~27 day components of sunspot area (Figure 2), cosmic ray flux (Figure 3 and Figure 8) and Melbourne and Central England temperature (Figure 5).

Comparison of Figure 5C and 7B show that in 2005 the cloud fraction and temperature vary in-phase at the beginning of the year, are out-of-phase in the middle of the year, and come back into phase towards the end of the year. The different phase relations during the year between cloud and temperature may be due to varying ratios of high and low cloud however no clear seasonal variation is evident. It is useful to assess what fraction of cloud variability is due to the ~27 day component. The mean cloud fraction and standard deviation of cloud fraction are for 1997, 0.37 and 0.24 and for 2005, 0.32 and 0.23. The standard deviations of the 27 day components of cloud fraction are for 1997, 0.025 and for 2005, 0.035 so the ~27 day component of cloud fraction amounts to 10% (1997) and 15% (2005) of the overall variance in cloud fraction. This percentage variation in cloud fraction leads to similar percentage variation in solar exposure or downward long wave radiation and, ultimately, to the significant variation, several degrees Centigrade, in surface temperature observed, see Figures 5B and 5C.

Figure 8A and 8B compare the ~27 day components of CRF and cloud fraction in Melbourne during 1997 and 2005.

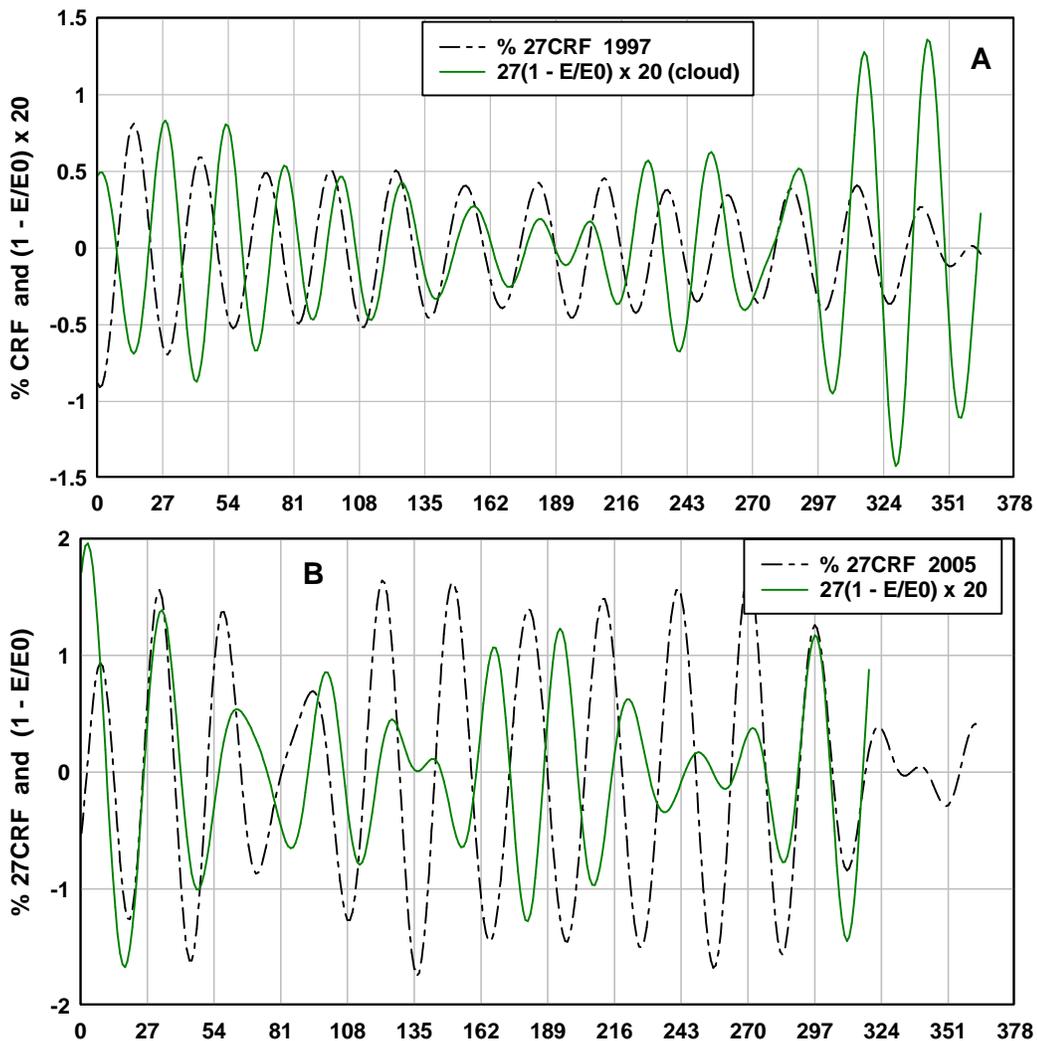

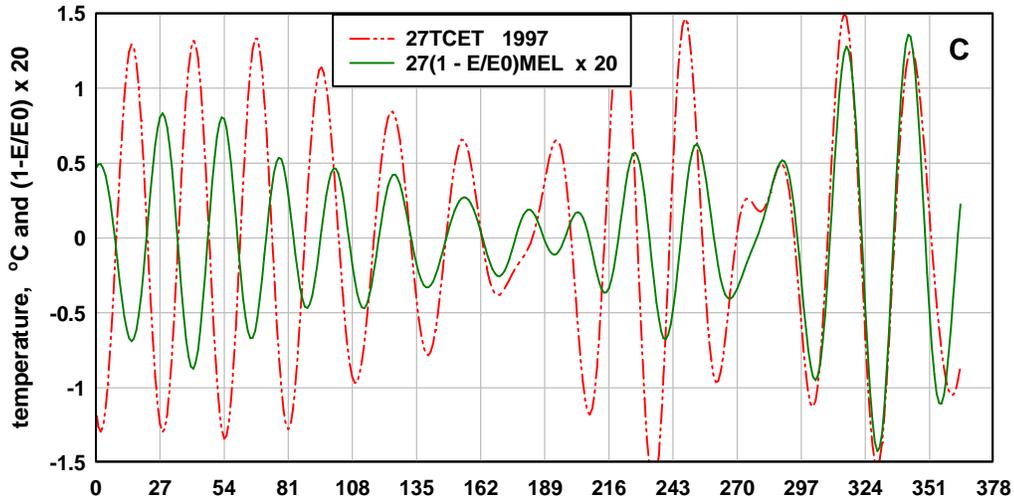

**Figure 8.** Compares the percentage variation in the ~27 day components of cosmic ray flux and cloud fraction in Melbourne for (A) 1997 and (B) 2005. In (C) we compare the variation of the ~27 day components of cloud fraction in Melbourne, 27(1 – E/E0)MEL and Central England temperature, 27TCET, during 1997.

Only broad conclusions can be drawn. In 1997, Figure 8A, CRF and cloud fraction are in anti-phase at the start of the year when, according to previous analysis, CRF variation is determined by a single equatorial coronal hole. The cosmic ray flux and temperature are mainly in-phase for the remainder of 1997. In 2005, Figure 8B, CRF and cloud fraction are mainly in-phase in summer and out-of-phase in winter. The 180° change of phase in 27CRF associated with the change of solar activity from one active longitude to the other occurs at day 81 and may be associated with the phase change in cloud fraction at that time. There may be a seasonal effect with cloud fraction mainly in phase with 27CRF in summer and out-of-phase in winter. The variation of the ~27 day components of TCET and cloud fraction in Melbourne is illustrated in Figure 8C. The apparent connection between the timing of cloud variation in Melbourne and the timing of temperature variation in England is indicative of a global influence. It is worth recalling that in the early part of 1997 the solar influence is dominated by the presence of an equatorial coronal hole, Harrison et al (2011), whereas in the last half of the year the solar influence appears to be dominated by two bursts of flares and sunspot area, Figure 1A and Zhang et al (2008).

Using solar exposure to estimate cloud cover is not conventional but a similar method has been used previously, Harrison et al (2008). Records of daily cloud cover, estimated in octal by visual observation, can be purchased from the Australian Bureau of Meteorology. Figures 9A and 9B compare the 27 day component of daily cloud cover for Hobart and 27(1 – E/E0) for Melbourne in 1997 and 2005. (Note that the cloud cover in octal has been reduced to cloud fraction by dividing by 8 octal). Hobart is about 550 km, 5 degrees latitude, to the south of Melbourne. However, in the main, the ~27 day components of cloud estimated by the two methods are similar. The exception is near the middle of 1997 when the components varied in anti-phase.

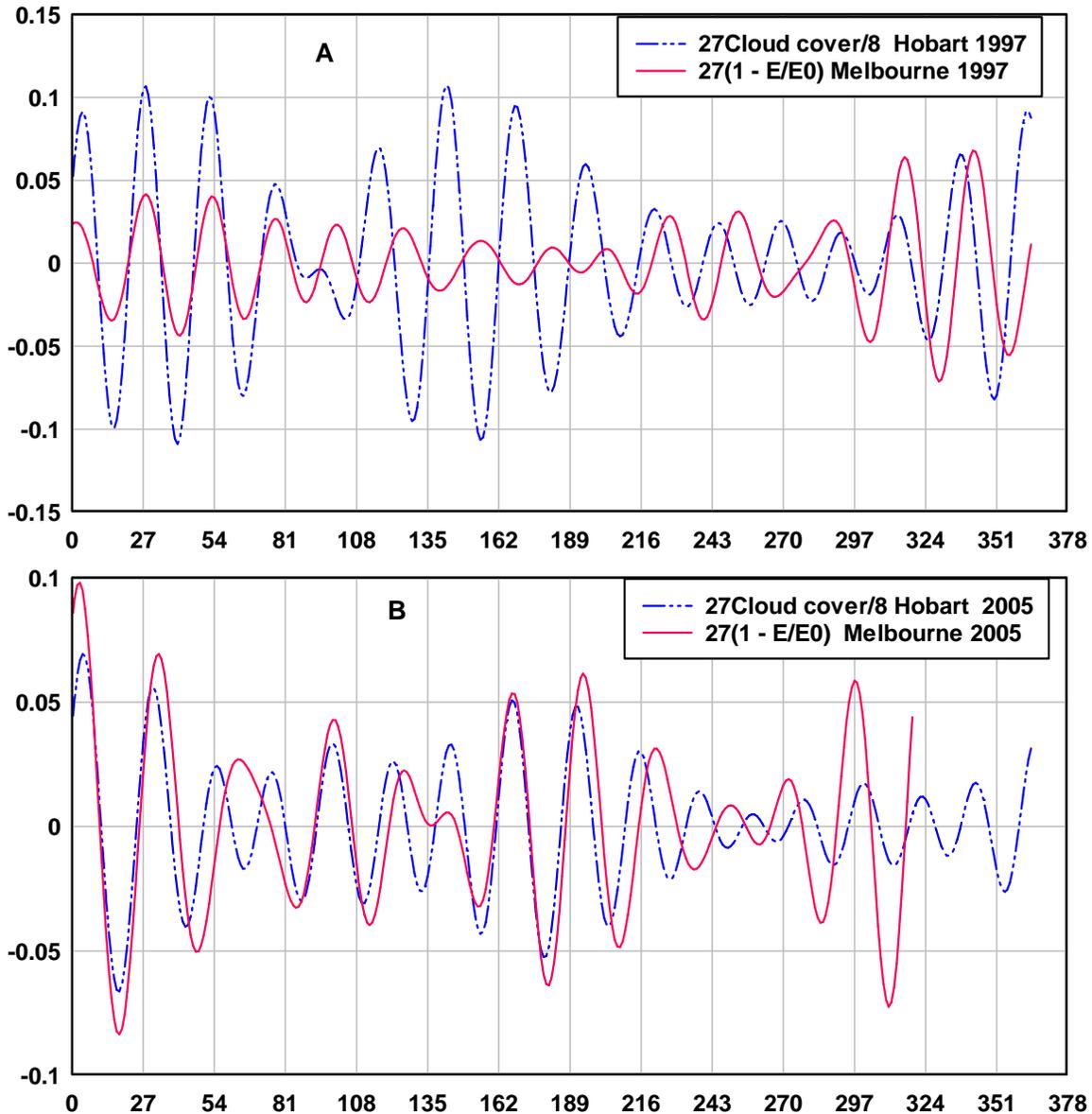

**Figure 9.** Compares ~27 day component of daily average cloud cover in Hobart obtained by visual observation with the ~27 component of daily average cloud fraction in Melbourne obtained from solar exposure measurements during (A) 1997 and (B) 2005.

### 7. Correlation with equatorial solar magnetic field.

In the extended time interval around Jan 01, 1997 the ~27 day components of the variables studied are phase-locked suggesting a simple connection. This appears to be a result of a long-lived coronal hole on the solar equator, Rouillard and Lockwood (2007). Harrison et al (2011) were able to show that the ~ 27 day components of cosmic ray flux and potential gradient were in-phase at this time and, in this paper we have shown that the ~27 day components of Central England temperature and Melbourne temperature vary near in-phase and the Melbourne cloud fraction varies out-of-phase with the ~27 day component of cosmic ray flux at this time, Figures 3A, 4A, 5B, 6, 8A, 8C and 9A. It would be interesting to know if a phase-locked response is a characteristic response to the occurrence of long–lived equatorial coronal holes. We use the term phase-locked advisably as we note that during the 2007 2008 coronal hole event Harrison et al (2013) observed cosmic ray flux and potential gradient varying in anti-phase in contrast to the 1996 1997 event when Harrison et al (2011) observed cosmic ray flux and potential gradient varying in-phase. Super-synoptic charts of the Suns magnetic field obtained at the Mount Wilson solar observatory for the years 1996 to 2001 have been published by Ulrich et al (2002). These show that sharply defined peaks in solar magnetic field on the solar equator occurred at ~27 day intervals during 1996 – 1997 and show, in the second half of 1999, seven similarly well defined peaks at ~27 day intervals, (see Figures 13 and 14, Ulrich et al, 2002). Figure 10 compares the ~27 day components of cosmic ray flux, Central England temperature, cloud fraction at Melbourne and Melbourne temperature in 1999. The cosmic ray flux and temperature components vary nearly in-phase for most of 1999 supporting the possibility that a single equatorial coronal hole is again influential. The vertical references lines mark times (taken from Figure 14, Ulrich et al (2002)) when the equatorial coronal hole was at the centre of the Suns disk.

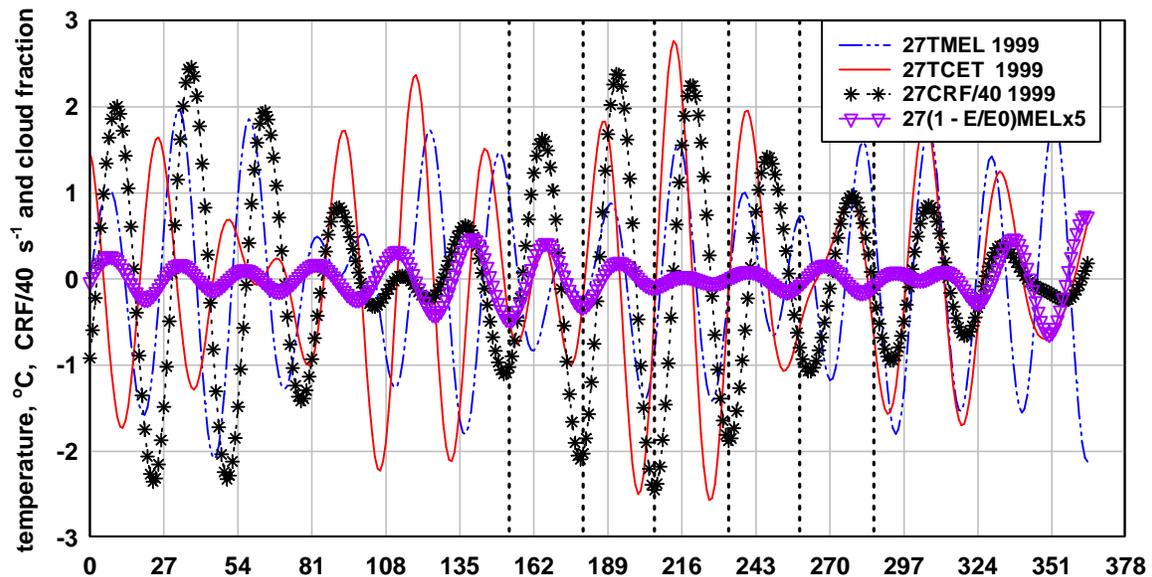

**Figure 10.** Compares ~27 day components of cosmic ray flux, cloud fraction , 27(1 – E/E0)MEL, and temperature in Melbourne and temperature in Central England for 1999. The dotted reference lines correspond to times when peaks in solar magnetic field due to an equatorial coronal hole were at the centre of the Sun's disc.

## 8. Estimation of significance of the observed correlations.

The conventional method of estimating significance of correlation is to compare observed correlation of the independent and dependent variables against correlation of the independent variable and red noise, Torrence and Compo (1998), Lockwood (2012). In the present work we are comparing band pass filtered components of solar activity and temperature. For example Figure 2C compares ~27 day components of sunspot area and temperature and Figure 5C compares ~27 day components of Central England temperature and Melbourne temperature. How significant are these correlations given that band pass filtering has a strong averaging effect?

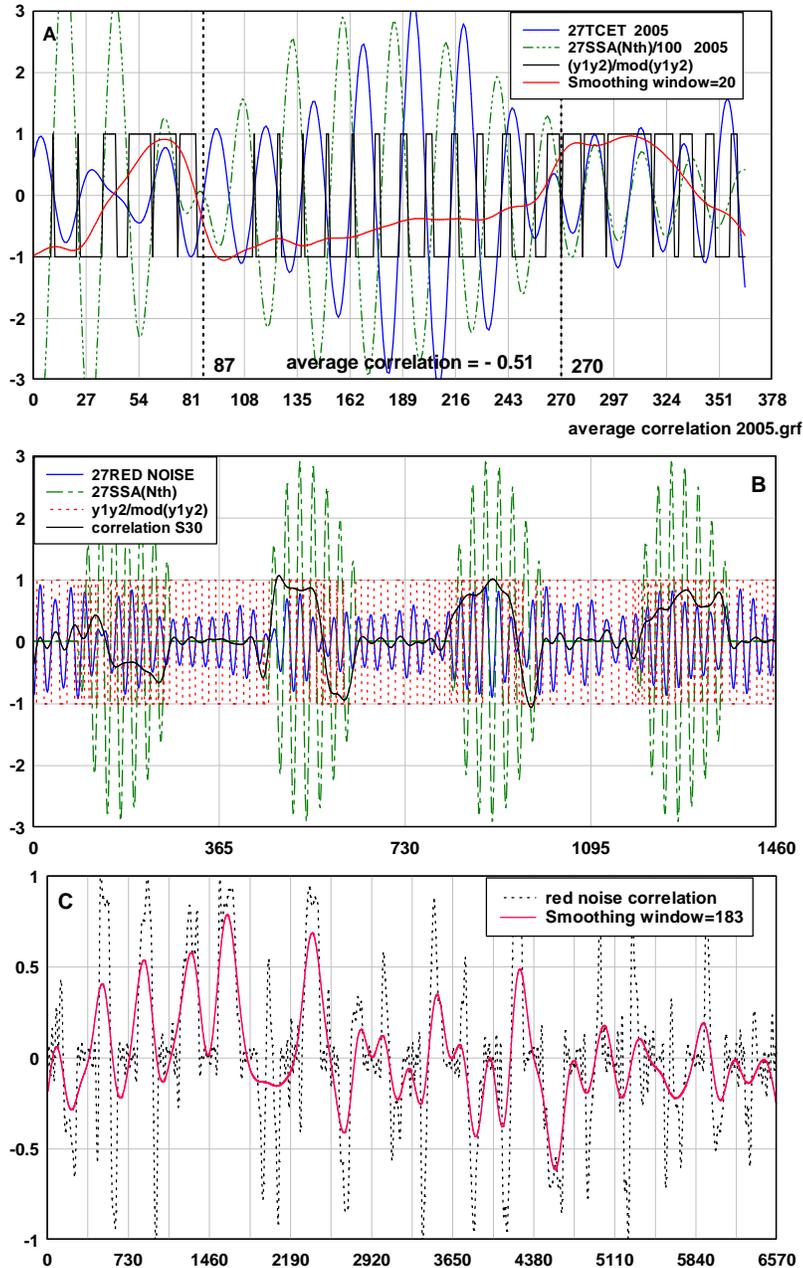

**Figure 11.** Illustrates the method of estimating significance of observed correlation. In (A) the correlation coefficient between 27SSA(Nth) and 27TCET during 2005 is estimated. In (B) a variation the same as the 27SSA(Nth) variation between days 87 and 270 is reproduced in a sequence of years, compared with red noise and the correlation coefficient estimated. In (C) a 183 day running average of the correlation coefficient is found so that the number of peaks above 0.5 or below -0.5 can be counted.

In this paper we postulate connection on the basis of similar amplitude variation and on phase correlation or phase anti-correlation over a number of solar cycles. It is difficult to test the significance of similar amplitude variation other than judgement by eye but it is relatively simple to estimate the significance or probability of phase correlation over several cycles. Figure 11A illustrates a simple test for in-phase or out-of -phase variation where we test the significance of the sunspot area and temperature correlation shown earlier, Figure 2C, which indicated an anti-phase correlation between temperature and sunspot area over seven solar cycles in the middle of the year 2005. The correlation coefficient of two variables y1 and y2 can be estimated from the running average of y1y2/mod(y1y2). This quantity is estimated for 27SSA(Nth) and 27TCET during 2005 and is shown as the red line in Figure 11A.

Over the seven cycles in the middle of 2005 the correlation has an average value of –0.51, representing moderate average negative correlation of the two variables in the half year interval (183 days) between the dotted reference lines. We can quantify how often positive or negative correlation with the observed seven cycle variation of sunspot area would occur if the temperature variation was unrelated red noise. Figure 11B compares, with red noise, an annual sequence of variations each one equal to the variation of sunspot area between the reference lines in Figure 11A. Also shown is the function y1y2/mod(y1y2) and the estimated correlation coefficient. By making a 183 day running average over the correlation coefficient in a long sequence, e.g. Figure 11C, we can find how often the average correlation exceeds 0.5 in magnitude by simply counting peaks above 0.5 or below -0.5. In a long sequence of 100 years, the probability of positive or negative correlation with red noise exceeding 0.5 in magnitude was 0.23. Thus, if the temperature variation in Figure 2C was unrelated red noise moderate positive or negative correlations, $> + 0.5$ or $< - 0.5$, would occur about once every four or five years i.e. quite frequently. However, in this work we are comparing temperature variations from more than one region with the solar activity variation. For example, Figure 2C and Figure 5 show that the 27 day components of temperatures in Melbourne and Central England are, respectively, positively and negatively correlated at magnitudes $> 0.5$ with 27SSA(Nth) during 2005. If the Melbourne temperature data and the Central England temperature data are both unrelated random red noise sequences the probability of correlation of both red noise sequences with the sunspot area variation exceeding 0.5 in magnitude is $(0.23)^2 = 0.05$ or about 5%. We conclude therefore that it is unlikely, at the 95% significance level, that the sunspot area and temperature variations observed in 2005 are unrelated.

**9. Correlations with temperature at Boulder, Colorado.**
The simple test of significance above indicates that the probability of correlations of sunspot area or cosmic ray flux with temperature being due to noise decreases as $(0.23)^n$ where n is the number of regions exhibiting strong, $> 0.5$ or $< -0.5$, correlation provided the regions where the temperature series are obtained are sufficiently separated that local weather effects at one location are unlikely to influence temperature at the other locations. Local weather effects at Melbourne are unlikely to influence weather in Central England. Another region distant from these two regions is Boulder, Colorado. Boulder, an inland, high altitude, location has a climate radically different from either Central England or Melbourne. In particular the average variability of temperature in Boulder is about two times higher in winter than in summer, whereas in Melbourne the variability is about two times higher in summer than in winter. In Central England, as remarked earlier, the average temperature variability is nearly constant over a year.

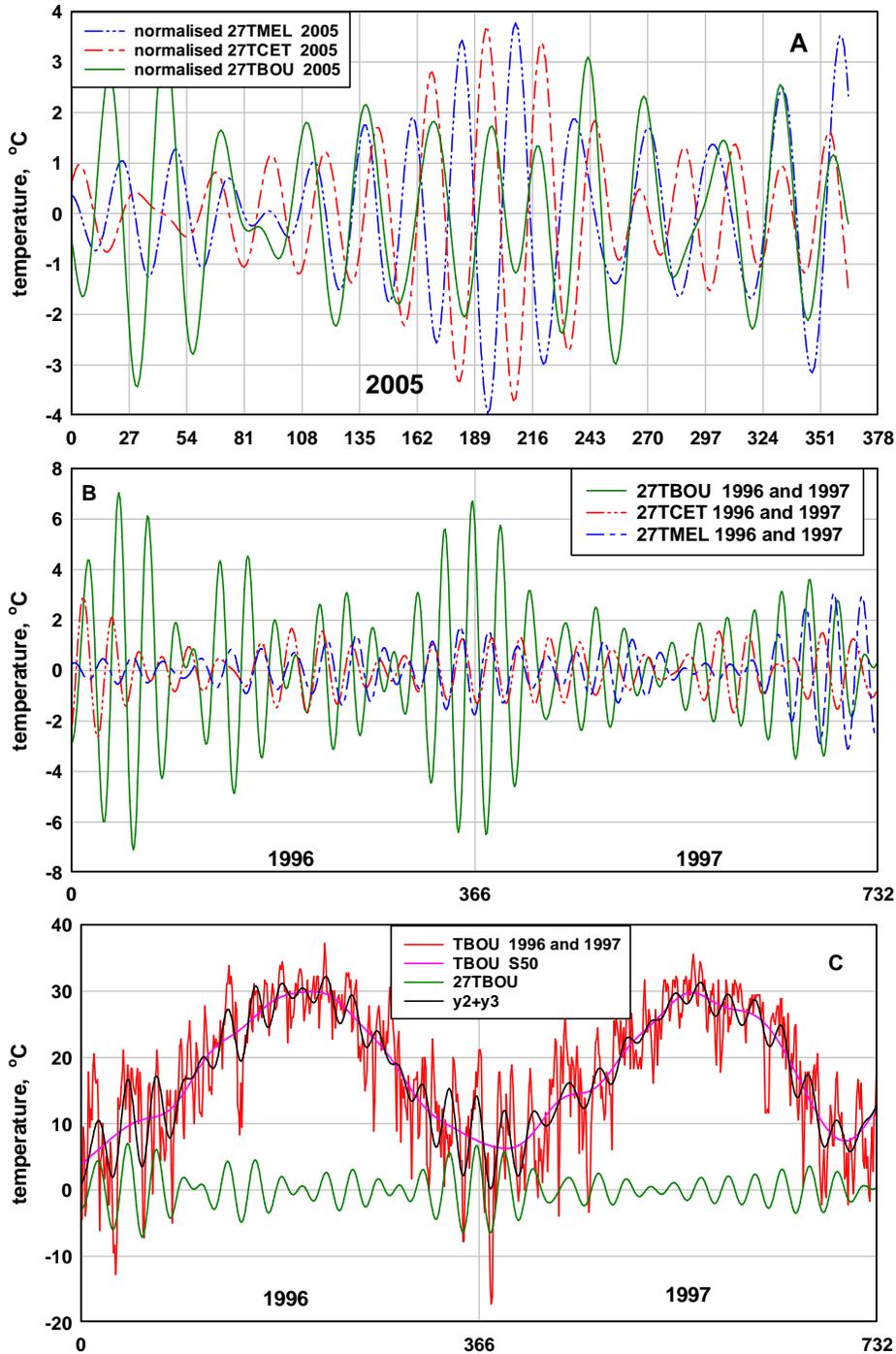

**Figure 12.** Compares ~27 components of temperature at Boulder, Central England and Melbourne during (A) 2005 and (B) 1996 1997. (C) compares the raw daily maximum temperature at Boulder during 1996 and 1997 with the ~27 day component.

Figure 12A compares the normalised, (normalised against average variability), temperature variations of 27TBOU, 27TCET and 27TMEL in 2005. In the middle of 2005 27TBOU varies in-phase with 27TCET and in anti-phase with 27TMEL. The phase relation of 27TBOU to the variation of sunspot area and cosmic ray flux can be seen by reference to Figures 2C and 3B respectively. The clear amplitude correlation that exists between 27TCET and 27TMEL is not seen. This may be due to the lack of cloud in Boulder over summer whereas Melbourne in winter is cloudy and Central England is always cloudy, even in summer. A conclusion that can be tentatively drawn is that when strong sunspot and flare activity, coronal mass ejections and Forbush decreases occur the ~27 day components of temperatures in the northern hemisphere vary in anti-phase with the ~27 day components of temperatures in the southern hemisphere.

Figure 12B compares 27TBOU, 27TCET and 27TMEL during 1996 and 1997. In the interval around Jan 01, 1997, 27TBOU varies in anti-phase with 27TCET and 27TMEL and in anti-phase with 27CRF, see Figure 6. A very tentative conclusion to be drawn is that the atmosphere may be responding in a modal manner to the ~27 day period variations in cosmic ray flux. For example, a $l = 3$, $m = 2$ spherical harmonic mode would match the observed phase of the temperature variations. Alternatively, the anti-phase variation of 27TBOU may be due to the much higher altitude of Boulder relative to either Central England or Melbourne. The amplitude of the 27TBOU variation is large during this interval, about 13$^o$C peak to peak and, overall, the amplitude of the 27TBOU variation is larger than either the 27TCET or 27TMEL variations. This may be due to the high altitude (1700 m) of Boulder and the increased cosmic ray flux there. Figure 12C shows the 27TBOU variation superimposed on the raw TBOU data for 1996 and 1997. It is evident that, around January 1997, the very large swings in the raw temperature are associated with the swings in 27TBOU and in 27CRF and, from Harrison et al (2011), swings in the ~27 day component of atmospheric potential gradient. This raises the possibility that extreme swings in temperature at various localities may often be associated with ~27 day swings in cosmic ray flux with, however, temperatures in different regions varying in-phase or anti-phase. It also raises the question as to how swings in cosmic ray flux amounting to a few percent can result in such large correlated or anti-correlated swings in temperature. The fact that the raw temperature swings are up to three times higher that the swings of the ~27 day component raises the possibility that harmonics of the ~27 day variations in cosmic ray flux i.e. variations at 13.5, 9 and 6.7 day periods may also be contributing. However, this possibility is not examined in this paper. There are two reasons: In the filtered ~27 day variations much of the short term synoptic weather noise is averaged out whereas in shorter term harmonic variations synoptic weather noise is likely to interfere more. Also there may be delays, of a few days, between cosmic ray variations and subsequent cloud and surface temperature variation (Svensmark et al 2009, Kirby 2007). Such delays would interfere more with the observation of shorter term correlation than with correlations of ~27 day variations. However, recent work, Nicoll and Harrison (2013), suggests that cosmic ray variations and cloud variations are almost simultaneous on the scale of ~27 day variations used in this study.

**10. Other large ~27 day temperature events correlated with solar activity.**
It was suggested in the previous section that large ~27 day swings in temperature in different regions may often be associated with ~27 day swings in cosmic ray flux or other forms of solar activity. There are other, large, ~27 day period temperature events apparently correlated with solar activity e.g. during 1927 and 1928, 1965 and 1966, 1983 and 1984, 2011 and 2012. In the 1983 and 1984 interval, around Jan 01 1984, Boulder experienced temperature swings of about 35°C of which the ~27 day component contributed swings of about 16°C, Figure 13A. (Note there is a short section of missing temperature data at the start of 1983 that has been replaced with smoothed average data). Figure 13B compares ~27 day components of sunspot area, cosmic ray flux and temperature in 1983 1984.

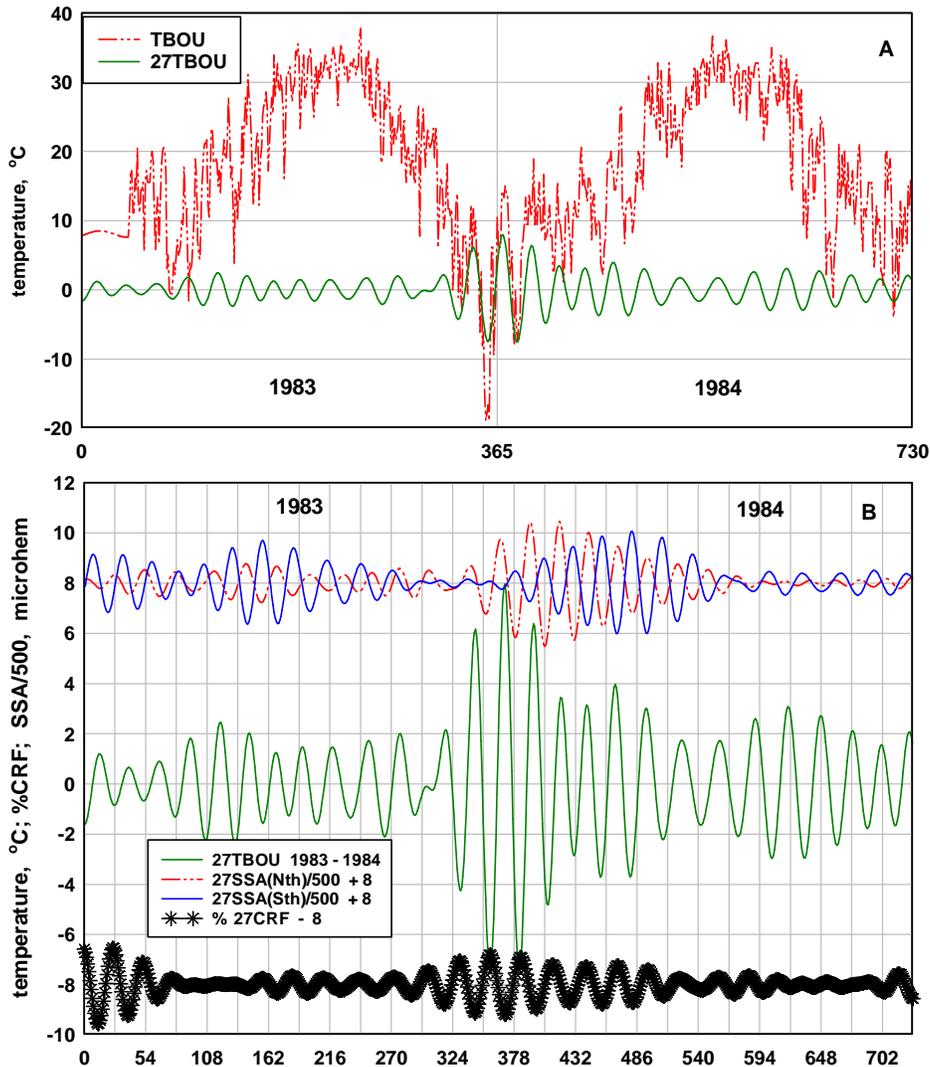

**Figure 13.** (A) Compares the raw daily maximum temperature at Boulder during 1983 and 1984 with the ~27 day component indicating that the ~27 day component contributed about half of the 35°C swing in temperature observed near Jan 01 1984. (B) shows that the swing in temperature around Jan 01 1984 was in-phase with the ~27 day component of cosmic ray flux and in anti-phase with the ~27 day component of sunspot area South.

On the basis of previous work in this paper we can draw very tentative conclusions about the complex relations illustrated in Figure 13B. First we note that the anti-phase variation of 27SSA(Nth) and 27SSA(Sth) indicates that there were active regions in the northern and southern hemispheres of the Sun on active longitudes separated by 180°. When sunspot activity occurred in the middle of 1983 the temperature responded in anti-phase to 27SSA(Sth). When that sunspot activity diminished we posit that an equatorial coronal hole emerged about day 270 resulting in the increase in the ~27 day component of cosmic ray flux at that time. By day 324 the variation in cosmic ray flux was causing strong variation in surface temperature with increases in cosmic ray flux resulting in decreases in temperature. Around day 378 sunspot activity in the southern solar hemisphere began to re-emerge and began again to influence the temperature which varied in anti-phase with 27SSA(Sth) until near the end of 1984. In the years 1927 and 1928 there is another interval with strong temperature swings correlated with solar activity, Figure 14. Clearly, over this entire interval there is a very complex situation and, in the absence of data on cosmic ray flux, we simply note that around Jan 01 1928 the ~27 day components of temperatures at Central England, Melbourne and Boulder were all swinging in anti-phase with 27SSA(Nth).

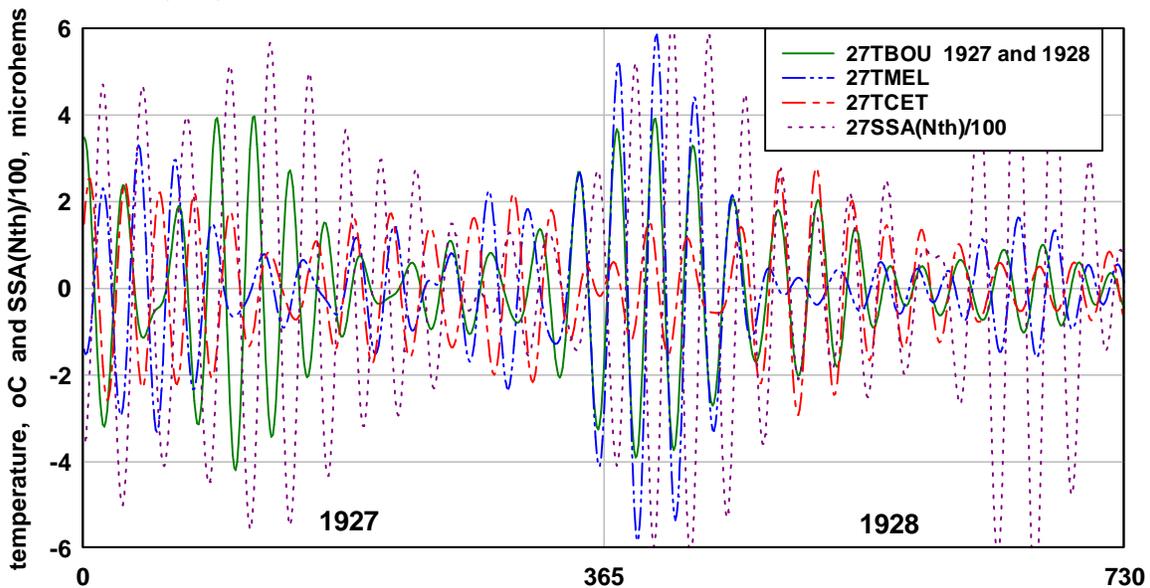

**Figure 14.** Shows that around January 1928 the ~27 day components of temperature at Boulder, Central England and Melbourne were all varying in anti-phase to the ~27 day component of sunspot area North.

**11. Discussion.**

The observations reported above indicate that when the connection solar activity – cosmic ray flux – cloud – surface temperature exists it is usually very complex. However, this is expected as all of the interconnections, solar activity – cosmic ray flux, cosmic ray flux – cloud and cloud – surface temperature are each, separately, complex and poorly understood. Some tentative conclusions concerning the phase and amplitude correlations observed were made in the previous sections. Here we make some general remarks. It is clear that large ~27 day variations in regional surface temperature are often correlated with ~ 27 day variation in solar activity. This conclusion differs from that drawn by Harrison et al (2013) who concluded that the correlation event between ~27 day variation of cosmic ray flux, cloud and temperature that they studied in 2007 2008 was very unusual. The present work observed persistent single equatorial coronal holes giving rise to persistent ~27 day variations in cosmic ray flux, cloud and temperature in several years and several regions. This type of variation sometimes overlapped with changes due to persistent regions of sunspot and flare activity. It appeared that cloud and temperature respond differently to the type of activity. When coronal holes are influential the phase relationship between the coronal hole, cosmic ray flux, cloud and temperature tends to be stable during the interval the equatorial coronal hole persists. However, there may be a $180^o$ phase difference in the response at different regions. When sunspot and flare activity is influential the phase between the ~27 day components of the variables appears to be less stable and may reflect the flip of solar activity from one active solar longitude to the other, Berdyugina (2005), Zhang et al (2008). While there is usually a stable phase relationship between the ~27 day components of variables during a specific event there does not seem to be a consistent phase relationship that applies to all events studied. The changing phase relationship from event to event is perhaps a reflection of a multi-process connection, Tinsley and Yu (2004), and the inconsistent phase relationship between ~27 day components of cloud and temperature from event to event, is perhaps a reflection of seasonal variation in cloud types and cloud altitudes. As mentioned previously in addition to the CRF - cloud mechanism there are other candidate solar influenced mechanisms that could lead to large regional temperature variation e.g. solar induced perturbations of the jet streams (Huth et al 2008, Woolings et al 2010 and Lockwood 2012) and variations in the jet streams due to stratospheric warming, Haigh (2003). A two mechanism process would go a long way to explaining the phase changes observed in this work. For example the phase changes between cosmic ray flux and Central England temperature in 1997 (Figure 3A) may reflect a second process associated with sunspot and flare activity becoming dominant over the cosmic ray flux – cloud mechanism. We introduce here a simple model to illustrate this possibility. Suppose the ~27 day variation of surface temperature, T, is the result of a mechanism determined by variation in cosmic ray flux, (CRF) and a second mechanism determined by variation in sunspot area (SSA). Assuming the two mechanisms work in opposition the resultant temperature variation can be expressed as T = aCRF – bSSA where a and b are constants. If CRF is the dominant influence as indicated in Figure 15A then a $180^o$ phase change in SSA due to the flipping (at the time indicated by the reference line) of activity from one active longitude to another would result in no phase shift in T relative to CRF. Alternatively if the influence of SSA is dominant, Figure 15B, the phase of the temperature variation would follow the phase shift in SSA. The possibility that the much of the complexity in the observed phase relationships can be explained by this type of two mechanism model will be examined in a later paper.

The frequent observation of large temperature variations correlated with cosmic ray flux variations around January e.g. 1997 (Figure 12B), 1999 (Figure 10), 1984 (Figure 13) and 1928 (Figure 14) is interesting and may be associated with closest approach of Earth to the Sun and maximum deviation of Earth from the ecliptic at this time. The correlation event in 2007 2008, Fig 4B and reported by Harrison et al (2013), was also centred on January 2008. Other large correlated temperature events occur in January in 1971, 1973, 1991 and 1996.

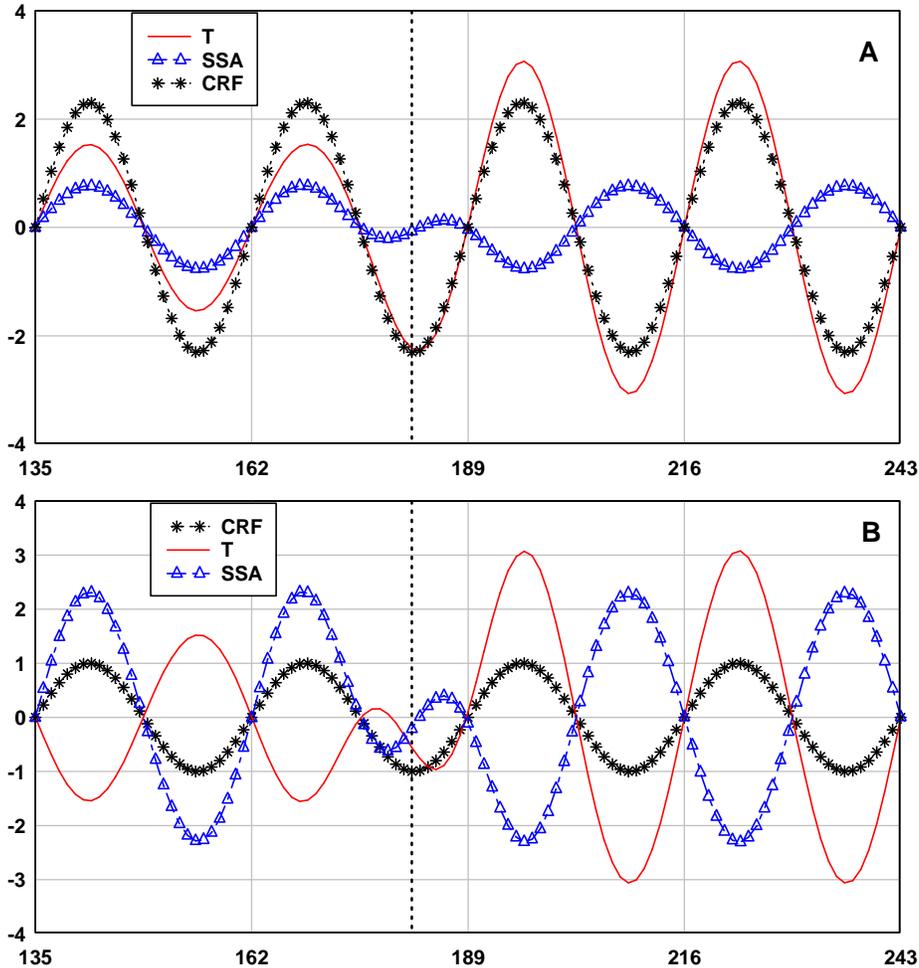

**Figure 15.** Illustrating a simple model to explain phase changes in ~27 day components relative to solar rotation.

This work compared the time variation of ~27 day components of solar activity with the time variation of ~27 day components of cloud and temperature. The objective was to select observations of sunspot and flare activity from prior, independent, work that exhibited clear, year long, time signatures of variation and establish whether the variation of cloud and temperature exhibited similar signatures of time variation in those years. This approach was largely successful in that similar amplitude signatures were observed. It was also found that cloud and temperature responded differently to activity associated with coronal holes and activity associated with sunspots and flares. When one form of activity occurred in isolation from the other the response of cloud and temperature to the solar activity was relatively simple. However, the cloud and temperature responses to solar activity were mainly either in-phase or out-of-phase in a manner that appeared to be regionally and seasonally dependent. One possibility is that the regional differences were due to a modal response of cloud and temperature to solar activity. When the two types of activity overlap, which clearly occurs most of the time, the cloud - temperature response is complex. Clearly much more work is required to understand even simple correlation events such as the 1996 1997, 1983 1984, 2005 and 2007 2008 events. However, the advantage of focusing on studying solar activity – cloud – temperature events in years that provide a depth of background solar and terrestrial information is clear. This work extends previous studies of the Sun-weather connection at single locations to studies across multiple regions providing evidence of the global nature of the connection and evidence that solar activity and temperature variations can be in-phase in one region and in anti-phase in another region.